\documentclass[preprint,3p,twocolumn,english]{elsarticle}

\usepackage{graphicx}
\usepackage{amssymb}

\journal{Physica E: Low Dimensional Systems and Nanostructures}

\begin{document}
\begin{frontmatter}

\title{Fano-Kondo Spin Filter}

\author{A. C. Seridonio$^{1,2}$, F. M. Souza\corref{A}$^{1,3}$, J. Del Nero$^{4,5}$ and I. A. Shelykh$^{1,6}$}

\address{$^{1}$International Center for Condensed Matter Physics,
Universidade de Bras\'{\i}lia, 04513, Bras\'{\i}lia, DF, Brazil\\
 $^{2}$Instituto de F\'{\i}sica, Universidade Federal Fluminense, 24310-346, Niter{\'o}i, RJ, Brazil\\
 $^{3}$Instituto de F\'isica, Universidade Federal de Uberl\^{a}ndia, 38400-902, Uberl\^{a}ndia, MG, Brazil\\
 $^{4}$Departamento de F\'isica, Universidade Federal do Par\'a, 66075-110, Bel\'em, PA, Brazil\\
 $^{5}$Instituto de F\'isica, Universidade Federal do Rio de Janeiro, 21941-972, Rio de Janeiro, RJ, Brazil\\
 $^{6}$Science Department, University of Iceland, Dunhaga 3, IS-107, Reykjavik, Iceland}
\cortext[A]{Corresponding Author: fmsouza@infis.ufu.br}

\begin{abstract}
We study spin-dependent conductance in a system composed of a
ferromagnetic (FM) Scanning Tunneling Microscope (STM) tip coupled
to a metallic host surface with an adatom. The Kondo resonance is
taken into account via the Doniach-Sunjic spectral function. For
short lateral tip-adatom distances and due to the interplay between
Kondo physics, quantum interfering effects and the ferromagnetism of
the tip, a spin-splitting of the Fano-Kondo line shape arises in the
conductance. A strong enhancement of the Fano-Kondo profile for the
majority spin component of the FM tip is observed. When the tip is
placed on the adatom, this gives a conductance 100\% polarized for a
particular range of bias voltage. The system thus can be used as a
powerful generator of spin polarized currents.
\end{abstract}

\end{frontmatter}

\section{Introduction}

One of the most fascinating phenomena due to strong correlated electrons
is the Kondo effect \cite{acH}. Observed originally in the context of magnetic
impurities diluted in bulk alloys, the Kondo physics results
in the enhancement of the material resistance for temperatures $T$
below the characteristic Kondo temperature $T_{K}$. In
the 90's, with the development of miniaturization techniques, systems
of quantum dots (QDs) coupled to a left and to a right electron reservoir
(e.g. metallic electrodes) became quite feasible \cite{um90}. In this
new context of QDs attached to electrodes, the Kondo effect
results in a wealth of novel features \cite{dgg98,smc98}. Probably,
the appearance in the dot density of states of a narrow peak (width
given by $k_{B}T_{K}$) at the Fermi level of the reservoirs is one
of the most remarkable feature. This peak gives rise to an enhancement
of the conductance, thus significantly improving the charge transport
in the system. In particular, in the presence of an external bias
voltage applied in a left and a right electrode (non-equilibrium regime),
two Kondo peaks appear pinned at the Fermi levels of the leads \cite{fs99,mk02}.
When the electrodes are ferromagnetic (FM), additional spin-dependent
effects emerge. For instance, due to a large local exchange field on
the QD, the Kondo resonance splits into two peaks, where
the energy separation can be tuned via the magnetization alignment
of the electrodes, their polarizations and a possible external magnetic
field \cite{jm01,jm02,msc04,yu05,rs06}. Nowadays, systems of a single
QD coupled to leads are routinely developed in the laboratories
and the observation of the Kondo resonance is frequent, including
the measurement of the Kondo peak splitting \cite{anp04,kh07}.

An alternative system is composed of a metallic host surface with
an adatom (adsorbed atom) and a FM tip \cite{krp07,acs08}.
Contrastingly to the well studied lead-QD-lead geometry, the present
system (tip-adatom-host) presents a more wealth physics in the sense
that quantum interference between distinct tunneling paths emerges.
The two possible interfering ways are the direct tunneling tip-host
and the indirect one, tip-adatom-host. In particular, below $T_{K}$
the Kondo resonance at the adatom spectral function provides the tunneling
channel through the adatom. The quantum interference mechanism results
in an asymmetric line shape of the Kondo resonance, that instead of
looks like a Lorentzian it becomes similar to a characteristic Fano
resonance \cite{Fano}. This Fano-Kondo resonance has been widely measured in experiments
with nonmagnetic tips \cite{vm98,nk02,kn02,pw05,experiment} and also discussed
in theoretical works \cite{as00,mp01,ou00}.

The development of FM tips \cite{sh00} makes the system tip-adatom-host
a potential candidate to future implementation of spintronic devices \cite{iz04}.
Recently, Patton \emph{et al.} \cite{krp07} studied theoretically
the effects of the tip ferromagnetism on the conductance in the Kondo
regime. It was found a splitting of the Fano line shape of the conductance.
Additionally, for FM tips it was investigated the effects of the lateral
tip-adatom separation on the conductance in the full range of the
Fano parameter, i.e., the rate between the tip-adatom and the tip-host
coupling strengths \cite{acs08}.

In the present work we calculate the spin-resolved conductance for
the system FM tip-adatom-host in the linear regime of tunneling. The
Kondo resonant peak is modeled via the Doniach-Sunjic spectral function \cite{DSfit}.
We predict a powerful generation of spin polarized currents even for
tips with a tiny magnetization.

The current can be full up or down polarized, depending
on the external bias. This effect comes from the spin splitting of
the Fano-Kondo resonance, that strongly suppresses one spin component
of the conductance in a particular bias range. The effects of the
lateral tip-adatom distance of this \emph{spin-polarizer device} are
also investigated.

\section{Theoretical Model}

We start from the model Hamiltonian

\begin{equation}
\mathcal{H}=\mathcal{H}_{SIAM}+\mathcal{H}_{tip}+\mathcal{H}_{T},\label{eq:1}\end{equation}
where the terms represent the host metal with the magnetic impurity
$(\mathcal{H}_{SIAM})$, the FM tip $(\mathcal{H}_{tip})$ and the coupling between them
$(\mathcal{H}_T)$. For the description of a magnetic impurity coupled to a host
metal, we use the Single Impurity Anderson Model (SIAM) \cite{PWA}

\begin{eqnarray}
\mathcal{H}_{SIAM} & = & \sum_{\vec{p}\sigma}\varepsilon_{\vec{p}}a_{\vec{p}\sigma}^{\dagger}a_{\vec{p}\sigma}+\varepsilon_{0}\sum_{\sigma}c_{0\sigma}^{\dagger}c_{0\sigma}\nonumber \\
 & + & Un_{0\uparrow}n_{0\downarrow}\nonumber \\
 & + &  v\sum_{\sigma}\left(f_{0\sigma}^{\dagger}c_{0\sigma}+H.C.\right),\label{eq:2}\end{eqnarray}
where $a_{\vec{p}\sigma}$ is the destruction operator for a conduction
electron with momentum $\vec{p}$ and spin $\sigma$, $c_{0\sigma}$ is
the destruction operator for an electron localized at the adatom site.
The parameters $\varepsilon_{0}$ and $U$ are the energy of the adatom level and the Coulomb
repulsion between two electrons with opposite spins at the adatom,  respectively.

For a sake of simplicity, we consider the case of a conduction band
with a flat density of states $\rho_{0}$ and half-width denoted by $D.$ In order to describe
the hybridization of this band with the adatom, we introduce a normalized
fermionic operator $f_{0\sigma}$ as follows

\begin{equation}
f_{0\sigma}=\frac{1}{\sqrt{N}}\sum_{\vec{p}}a_{\vec{p}\sigma},\label{eq:3}\end{equation}
where $N$ denotes the total number of the conduction states.

\begin{figure}
\includegraphics[%
    width=1.0\columnwidth,
    height=0.70\linewidth,
    keepaspectratio]{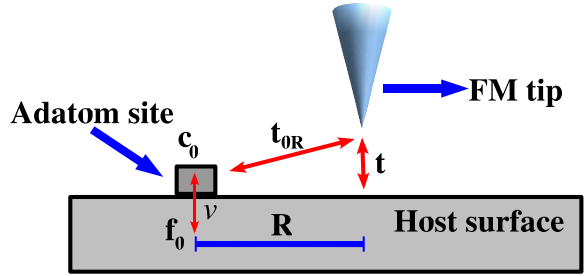}

\caption{\label{fig1}Scanning Tunneling Microscope (STM) device with a ferromagnetic
(FM) tip with an adsorbed magnetic atom (Kondo adatom) on the surface of a normal
metal (Host surface). In the Host surface, the adatom site $(c_{0})$
is hybridized with the metal conduction band $(f_{0})$ through the
hopping $v$. The interference between the hopping elements
$t_{0R}$ and $t$ leads to the Fano-Kondo profile in the conductance.}
\end{figure}

As we are focusing on the Kondo regime, we consider only the range of
system parameters favoring the formation of a localized magnetic
moment at the adatom, namely
$\varepsilon_{0}<\varepsilon_{F},\varepsilon_{0}+U>\varepsilon_{F}$
and $\Gamma\ll|\varepsilon_{0}|$, $\varepsilon_{0}+U$. For
temperatures $T \ll T_K,$ this magnetic moment becomes screened by
the host conduction electrons, due to the antiferromagnetic correlations and
the Kondo peak with half-width $k_{B}T_{K}$ and centered at the Fermi
energy, appears in the adatom spectral density. Here we use the
same estimate for the Kondo temperature as in \cite{CHZ},
\begin{equation}
k_{B}T_{K} = \sqrt{\frac{\Gamma U}{2}} \textrm{exp} [\pi \varepsilon_{0} (\varepsilon_{0}+U)/2 \Gamma U].\label{eq:Kondo_TK}\end{equation}

In this work the electron-electron interactions inside the FM tip are
neglected and thus

\begin{equation}
\mathcal{H}_{tip}=\sum_{\vec{k}\sigma}(\varepsilon_{\vec{k}}+eV)b_{\vec{k}\sigma}^{\dagger}b_{\vec{k}\sigma},\label{eq:4}\end{equation}
where  $b_{\vec{k}\sigma}$ is the destruction operator for a tip conduction
electron and $V$ is the tip bias voltage. As we consider the case of a
ferromagnetic tip, the corresponding density of states becomes
spin-dependent and is described by

\begin{equation} \rho_{tip}^{\sigma}=\rho_{0}\left(1+\sigma
p\right),\label{eq:tip_DOS}\end{equation} with $\sigma=\pm1$
denoting the spin orientations, $p$ being the tip magnetization.

The tunneling Hamiltonian couples the tip with the host metal and the
adatom, i.e.,

\begin{eqnarray}
\mathcal{H}_{T} & = & \sum_{\vec{k}\sigma}b_{\vec{k}\sigma}^{\dagger}\left(t\sum_{\vec{p}}\varphi_{\vec{p}}(\vec{R})a_{\vec{p}\sigma}+t_{0R}c_{0\sigma}\right) \nonumber \\
 & + & H.C.\label{eq:5}\end{eqnarray}
where the terms  $t$  and $t_{0R}=t_{0}\textrm{exp}(-k_{F}R)$ correspond to
the tip-host and tip-adatom hopping terms, respectively. Following \cite{mp01},
we introduce the exponential decay of $t_{0R}$ with increasing of
the tip-adatom lateral distance $R$. The amplitude of the tip-host metal
tunneling depends on the value of the conduction band wave function
$\varphi_{\vec{p}}(\vec{R})$ calculated at the tip position.

\section{The Conductance Formula}

The derivation  of the conductance formula treats the tunneling
Hamiltonian (\ref{eq:5}) as a linear perturbation \cite{as00}. As we consider
$eV\ll D$ and $T\ll T_{K}$ in the Kondo regime, we can safely evaluate the conductance formula at $T=0,$
which results in

\begin{equation}
G(eV,T\ll T_{K},R)=\sum_{\sigma}G_{0}\left[\frac{\rho_{tip}^{\sigma}}{\rho_{0}}\right]\left[\frac{\rho_{0R}^{\sigma}}{\rho_{0}}\right],\label{eq:conductance}\end{equation}
where
\begin{equation}
G_{0}=\left(2\pi t \rho_{0}\right)^{2}\left(\frac{e^{2}}{h}\right)\label{eq:G_max}\end{equation}
is the background conductance and

\begin{equation}
\rho_{0R}^{\sigma}(eV)=\rho_{0}\left\{ 1-\left(\frac{\mathcal{F}_{k_{F}R}}{\rho_{0}}\right)^{2}+\frac{\tilde{\rho}_{0R}^{\sigma}(eV)}{\rho_{0}}\right\} \label{eq:rho_s2}\end{equation}
is the host metal density of states in the presence of the Kondo adatom with the tip.
The function

\begin{equation}
\mathcal{F}_{k_{F}R}=\sum_{\vec{k}}\varphi_{\vec{k}}(\vec{R})\delta(\varepsilon_{k}-\varepsilon_{F})=\rho_{0}J_{0}\left(k_{F}R\right)\label{eq:spatial}\end{equation}
with $J_0$ being the zeroth-order Bessel function, accounts for the spatial
dependence of the conductance on the lateral tip-adatom distance
$\vec{R}$. The density in the spectral representation over the
eigenstates of the Hamiltonian (\ref{eq:2}) given by expression

\begin{eqnarray}
\tilde{\rho}_{0R}^{\sigma}(eV) & = & \sum_{m}|\left\langle m\right|\left(\frac{\mathcal{F}_{k_{F}R}}{\rho_{0}}\right)f_{0\sigma}\nonumber \\
 & + & {\pi\rho_{0}v q_{R}c_{0\sigma}}\left|{\Omega}\right\rangle |^{2}\nonumber \\
 & \times & \delta(E_{m}-E_{\Omega}-eV),\label{eq:rho_0}\end{eqnarray}
accounts for the quantum interference between the tunneling paths
$t_{0R}$ and $t,$ characterized by the Fano factor \cite{mp01}

\begin{equation}
q_{R}=\left(\pi\rho_{0}v\right)^{-1}\left(\frac{t_{0R}}{t}\right)=q_{R=0}\textrm{exp}(-k_{F}R).\label{eq:Fano_1}\end{equation}
The cases $(q_{R}\gg 1)$ and $(q_{R}\ll 1)$
correspond to the domination of tip-adatom and tip-host metal
tunneling, respectively, while the limit $(q_{R}\approx1)$
corresponds to the case where these two processes are of comparable
intensity. After some algebra (see Ref. \cite{acs08} for relevant
details), we can write Eq.(\ref{eq:rho_s2}) as

\begin{eqnarray}
\rho_{0R}^{\sigma}(eV) & = & \rho_{0}\left\{ 1+\left[\left(\frac{\mathcal{F}_{k_{F}R}}{\rho_{0}}\right)^{2}-\left(\tilde{q}_{R}\right)^{2}\right]\right.\nonumber \\
 & \times & \pi \Gamma\rho_{ad\sigma}(eV) \nonumber \\
 & + & \left.2\Gamma\left(\frac{\mathcal{F}_{k_{F}R}}{\rho_{0}}\right)\tilde{q}_{R}\Re\left\{ G_{ad\sigma}^{Ret}(eV)\right\} \right\}\nonumber \\
  \label{eq:rho_0_b}\end{eqnarray}
where

\begin{equation}
\tilde{q}_{R}=q_{R}+\left(\frac{\mathcal{F}_{k_{F}R}}{\rho_{0}}\right)q\label{eq:Fano_effective}\end{equation}
is the effective Fano factor,

\begin{equation}
q=\frac{\sum_{\vec{k}}\frac{1}{eV-\varepsilon_{\vec{k}}}}{\pi\rho_{0}}=\frac{1}{\pi}\ln\left|\frac{eV+D}{eV-D}\right|\label{eq:Fano_2}\end{equation}
is the Fano factor for the SIAM in the absence of the tip \cite{Fano,mp01} and

\begin{eqnarray}
\rho_{ad}^{\sigma}(eV) & = & -\frac{1}{\pi}\Im\left\{ G_{ad\sigma}^{Ret}(eV)\right\} \nonumber \\
 & = & \frac{1}{\pi\Gamma}\Re\left[\frac{i\Gamma_{K}}{(eV+\sigma\delta)+i\Gamma_{K}}\right]^{\frac{1}{2}}\nonumber \\
 & = & \frac{1}{\pi\Gamma}\sin^{2}\delta_{eV}^{\sigma}\label{eq:DOS_adatom}\end{eqnarray}
is the adatom spectral density determined by the interacting adatom
Green function $G_{ad\sigma}^{Ret}(eV),$ written in terms of
the Doniach-Sunjic formula \cite{DSfit} and the phase shift
$\delta_{eV}^{\sigma}$ of the conduction states, respectively.

The Kondo peak splitting $\delta$ originates from the ferromagnetic
exchange interaction between the tip and the magnetic adatom
\cite{jm01,yu05}, resulting in the lifting of the spin degeneracy for the
adatom level $\varepsilon_{0}$. Following the ``poor man's'' scaling
method \cite{jm01,krp07}, this splitting can be estimated as

\begin{eqnarray}
\delta & = & \varepsilon_{0 \downarrow}-\varepsilon_{0 \uparrow}\nonumber\\
 & = & \frac{\gamma_{tip}^{\uparrow}+\gamma_{tip}^{\downarrow}}{2\pi}p\ln(D/U)\nonumber \\
 & = & \left[\frac{\gamma_{tip}}{\pi}\ln(D/U)\right]p\exp\left(-2k_{F}R\right)\label{eq:Kondo_splitting}\end{eqnarray}
where

\begin{eqnarray}
\gamma_{tip}^{\sigma} & = & \pi\rho_{tip}^{\sigma}\left|t_{0R}\right|^{2} \nonumber \\
& = & \gamma_{tip}\left(1+\sigma p\right)\exp\left(-2k_{F}R\right)\label{eq:gama_tip_sigma}.\end{eqnarray}

In the regime $eV \ll D$ one can neglect the contribution of
Eq.(\ref{eq:Fano_2}) in Eq.(\ref{eq:Fano_effective}), thus we prefer to express the spin-resolved
conductance as

\begin{equation}
G^{\sigma}/G_{max}=\left[\frac{\rho_{tip}^{\sigma}}{\rho_{0}}\right]\left[\frac{\rho^{\sigma}(eV)}{\rho_{0}}\right],\label{eq:spin_resolved}\end{equation}
where

\begin{equation}
G_{max}=\left(1+q_{R}^{2}\right)G_{0}\label{eq:Gmax_2}\end{equation}
and

\begin{equation}
\rho^{\sigma}(eV)=\rho_{0R}^{\sigma}(eV)/\left(1+q_{R}^{2}\right),\label{eq:LDOS}\end{equation}
which is a function bounded by two in units of $G_{max}.$ Using the expressions

\begin{equation}
\tan\delta_{eV}^{\sigma}=-\frac{\Im\left\{
G_{ad\sigma}^{Ret}(eV)\right\} }{\Re\left\{
G_{ad\sigma}^{Ret}(eV)\right\} },\tan\delta_{q_{R}}=q_{R}
\label{eq:delta_ev_sigma}\end{equation}
and particularizing for $R=0,$ Eq.(\ref{eq:spin_resolved}) reduces to the simple form

\begin{equation}
G^{\sigma}/G_{max}=\left(1+\sigma
p\right)\cos^{2}\left(\delta_{eV}^{\sigma}-\delta_{q_{R=0}{}}\right).\label{eq:G_R_0}\end{equation}
Finally, the spin polarization of the tunneling current is determined
as
\begin{equation}
\wp=\frac{G^{\uparrow}-G^{\downarrow}}{G^{\uparrow}+G^{\downarrow}}.\label{eq:spin_polarization}\end{equation}

\section{Results}

For quantitative analysis of the conductance we choose a following
set of system parameters: $q_{R=0}=1,$ $\varepsilon_{0}=-0.9$eV,
$\gamma_{tip}=\Gamma=0.2$eV, $U=2.9$eV, $D=5.5$eV, $T_{K}=50$K and
$k_{F}=0.189$\AA{}$^{-1}$ \cite{krp07,ou00}. The Kondo temperature
was estimated using Eq.(\ref{eq:Kondo_TK}).

\begin{figure}
\includegraphics[%
    width=2.0\columnwidth,
    height=0.7\linewidth,
    keepaspectratio]{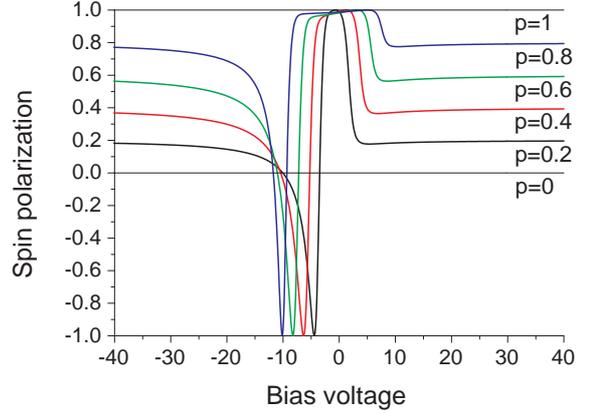}

\caption{\label{fig2}Spin Polarization at
$R=0$ ($\wp$) against bias voltage $eV$
in units of the Kondo half-width $\Gamma_{K}$ for $q_{R=0}=1$ and differing tip magnetization parameter $p.$  Starting at $p=0$ we find zero polarization for the conductance.
For nonzero $p$'s, the $\wp$ presents a maximum value equal one
$(\wp=+1)$, which is approximately constant for a certain bias range
(plateau). We also find a dip which gives $\wp=-1$. The parameters
adopted here follow Refs. \cite{krp07} and \cite{ou00}, i.e., $\varepsilon_{0}=-0.9$eV,
$\gamma_{tip}=\Gamma=0.2$eV, $U=2.9$eV, $D=5.5$eV, $T_{K}=50$K, and $k_{F}=0.189$\AA{}$^{-1}$.}
\end{figure}

Figure (\ref{fig2}) shows the Spin Polarization  $\wp$ as a function
of the bias voltage for differing tip magnetization parameter $p$.
The case $p=0$ gives no spin polarization as it is indeed expected
for a nonmagnetic lead. When $p\neq0$, $\wp$ reveals two remarkable
features: (i) it attains a maximum plateau ($\wp=+1$) around $eV=0$,
and (ii) it shows a narrow dip at some specific negative bias
voltage, yielding to $\wp=-1$. Increasing $p$, it is possible to
enlarge the plateau in which $\wp=+1$, and the dip ($\wp=-1$) moves
toward negative biases. Particularly, for $p=1$ (full polarized tip)
the spin polarization loses its structure, assuming a constant value
$\wp=1$. Note however, that even for small tip polarizations (e.g.
$p=0.2$), it is still possible to find $\wp=\pm1$, around $eV=0$. So
this system operates as a powerful \emph{current polarizer}, even
for tips with weak magnetization (small $p$). For relatively high
voltages, the Spin Polarization reaches a constant value equal to
the tip magnetization parameter $p$. This system, thus, provides the
means to choose the spin polarization from $-1$ to $+1$, by simple
tuning a tip's voltage $V$. A variety of systems capable to tune the
current polarization such as  spin-filters
\cite{jce98,Frustaglia,KimKiselev,ShelykhSplitter} and the
spin-diodes \cite{fms07} have emerged in the context of spintronics.
Some of them were recently realized experimentally \cite{cam08}.
It is also possible to control dynamically the current polarization by using
time dependent bias voltages \cite{fms07_2,fms09}.
However, to the best of our knowledge, the system we consider here
is first to use the interplay between Kondo effect and quantum
interference to control and amplify a current polarization.

\begin{figure}
\includegraphics[
    width=2.0\columnwidth,
    height=0.63\linewidth,
    keepaspectratio]{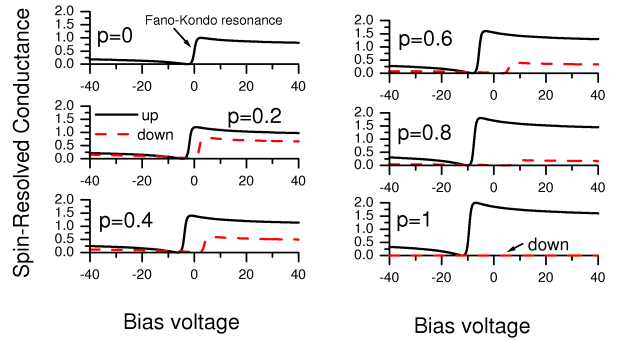}

\caption{\label{fig3}Spin-Resolved Conductances at
$R=0$ (Eq.(\ref{eq:G_R_0})) as a function of the bias
voltage $eV$ in units of the Kondo half-width $\Gamma_{K}$ for $q_{R=0}=1.$ For a normal tip ($p=0$) the spin
resolved conductances, $G^{\uparrow}$ and $G^{\downarrow}$, are
degenerate and present the typical Fano-Kondo line shape. For ferromagnetic
tip ($p\neq0$) the Fano-Kondo profile is spin-splitted, thus resulting
in a region in which $G^{\downarrow}\approx0$ while $G^{\uparrow}$
is maximum. In this region the conductance spin-polarization $\wp$
develops a plateau with $\wp=1$ (full polarized transport). Parameters
as in Fig. (\ref{fig2}).}
\end{figure}

Figure (\ref{fig3}) shows separately the Spin-Resolved Conductances,
$G^{\uparrow}$ and $G^{\downarrow}$. For $p=0$ we find
$G^{\uparrow}=G^{\downarrow}$, with Kondo resonance having
asymmetric Fano line shape coming from the interference between the
two possible tunnelings paths. Similar Fano-Kondo profile has been
observed in systems with normal metal tips
\cite{vm98,nk02,kn02,pw05,experiment}. With increasing of $p$, the
conductances demonstrates a spin-splitting with $G^{\uparrow}$
moving to the left and $G^{\downarrow}$ to the right, thus
generating a \emph{window} of completely spin polarized transport
($G^{\downarrow}\approx0$). This window corresponds to $\wp=1$
plateau in Fig. (\ref{fig2}). Additionally, the $\wp=-1$ dip in Fig.
(\ref{fig2}) comes from the minimum of $G^{\uparrow}$ in the
negative voltage range in Fig. (\ref{fig3}). While $G^{\uparrow}$
increases in amplitude, $G^{\downarrow}$ is suppressed. This happens
because $G^{\sigma}$ is weighted by the tip density of states, which
is proportional to $(1\pm p)$ (see Eq.(\ref{eq:G_R_0})). In the
limiting case $p=1$ only $G^{\uparrow}$ remains.

\begin{figure}
\includegraphics[%
 width=2.0\columnwidth,
 height=0.70\linewidth,keepaspectratio]{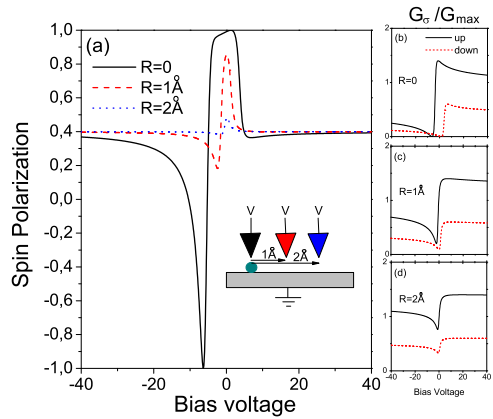}

\caption{\label{fig4}(a) Spin polarization $\wp$ against bias voltage $eV$ in units of the Kondo half-width $\Gamma_{K}$
for fixed tip magnetization $p=0.4$, $q_{R=0}=1$ and differing
tip-adatom lateral displacements $R$. When $R$ increases the spin
polarization structure (maximum plateau and dip) is washed out and
the overall curve tends to a background polarization $p$. In panels
(b)-(d) we present the evolution of the spin resolved conductances,
$G^{\uparrow}$ and $G^{\downarrow}$, with the distance. Increasing
$R$ the splitting of $G^{\uparrow}$ and $G^{\downarrow}$ is quenched
and the asymmetric Fano resonances (for each spin) evolve toward a
more symmetric profile. In the inset we sketch the tip displaced from
the adatom for $R=1$\AA{} and 2\AA{}. A voltage $V$ is applied
on the tip. Others parameters as in Fig. (\ref{fig2}).}
\end{figure}

Finally, in figure (\ref{fig4}) we show the effects of the lateral
tip-adatom separation on the Spin Polarization $\wp$. As $R$
increases, the minimal and maximal values of $\wp$ are reduced, and
the whole curve approaches a flat line corresponding to a
magnetization of the tip  $p$. This behavior of $\wp$ can be
understood from right panels (b)-(d), where spin-resolved conductances
$G^{\sigma}$ are shown. Increasing $R$, the Fano-Kondo resonances
for both $G^{\uparrow}$ and $G^{\downarrow}$ tend to a more
symmetric antiresonance and the spin-splitting is quenched. It is
valid to note that similar evolution of the Fano-Kondo line shape
with $R$ was experimentally reported in Ref. \cite{vm98} for an
unpolarized tip. The overall evolution picture in this experiment
consisted of a Fano like resonance for $R\approx0$, a more symmetric
antiresonance for intermediate $R$ and an approximately flat profile
(background) for large enough $R$. This general picture is also seen
for each $G^{\sigma}$ in the present spin-dependent case. However,
due to the ferromagnetism of the tip, $G^{\uparrow}$ remains above
$G^{\downarrow}$ which makes the spin-polarization $\wp$ approach
the background polarization $p$.

\section{Conclusions}

We have calculated Spin-Resolved Conductances for a system with a
ferromagnetic tip coupled to an adatom on a nonmagnetic
metallic surface (host). The conductance line shape is characterized
by a Fano resonance, which differs for each spin component. In particular,
we find a bias range in which $G^{\downarrow}\approx0$ while $G^{\uparrow}$
is maximum, due to the Fano interference between tunneling paths.
This contrasting features for $G^{\uparrow}$ and $G^{\downarrow}$
develops a plateau in the Spin Polarization  $\wp$ in which
the current can be full polarized ($\wp=+1$). By increasing the tip
magnetization parameter $p$, it is possible to enlarge this plateau,
thus covering a more broaden bias window. A full down polarized current
($\wp=-1$) can also be achieved by simple tuning the bias voltage.
This system thus operates as a powerful source of polarized current,
with its polarization being dependent on the biases.

\section{Acknowledgments}

This work was supported by the Brazilian agencies IBEM, CAPES, FAPESPA and FAPERJ.


\begin{thebibliography}{10}
\bibitem{DSfit}The Doniach-Sunjic formula was recently used to fit experimental Kondo
resonances, revealing a quite good agreement. See \cite{experiment}.
\end{thebibliography}
\end{document}